\documentclass[%
 reprint,
 superscriptaddress,
 amsmath,amssymb,
 aps,
prb,
]{revtex4-2}

\usepackage{xfrac}
\usepackage{graphicx}
\usepackage{dcolumn}
\usepackage{bm}
\usepackage{amssymb}
\usepackage{amsthm}
\usepackage[version=4]{mhchem}
\usepackage{subcaption} 
\usepackage{float} 
\usepackage{xcolor}
\usepackage{comment}
\usepackage{newunicodechar}
\usepackage{soulutf8}
\setstcolor{red}   

\newunicodechar{−}{\ensuremath{-}}

\begin{document}
\preprint{APS/123-QED}

\title{Electronic and magnetic properties of light rare-earth cubic Laves compounds derived from XMCD experiments}

\author{Vilde G. S. Lunde}\thanks{Contact author: vilde.lunde@ife.no}
\affiliation{Department for Hydrogen Technology, Institute for Energy Technology, PO Box 40, NO-2027, Kjeller, Norway}
\author{Benedicte S. Ofstad}
\affiliation{Department for Hydrogen Technology, Institute for Energy Technology, PO Box 40, NO-2027, Kjeller, Norway}
\author{{\O}ystein S. Fjellv{\aa}g}
\affiliation{Department for Hydrogen Technology, Institute for Energy Technology, PO Box 40, NO-2027, Kjeller, Norway}
\author{Philippe Ohresser}
\affiliation{Synchrotron SOLEIL, L'Orme des Merisiers, 91190, Saint-Aubin, France}
\author{Anja O. Sj{\aa}stad}
\affiliation{Chemistry Department and Center for Material Science and Nanotechnology, University of Oslo, NO-0315, Norway}
\author{Bj{\o}rn C. Hauback}
\affiliation{Department for Hydrogen Technology, Institute for Energy Technology, PO Box 40, NO-2027, Kjeller, Norway}
\author{Christoph Frommen} \thanks{Contact author: christoph.frommen@ife.no}
\affiliation{Department for Hydrogen Technology, Institute for Energy Technology, PO Box 40, NO-2027, Kjeller, Norway}

\date{\today}

\begin{abstract}

This work presents electronic and magnetic properties of selected members in the cubic Laves phase series \ce{Nd_{1-x}Pr_xCoNi} (0~$\leq$~x~$\leq$~1) and \ce{Ce_{0.25}Pr_{0.75}CoNi}, together with the corresponding binary compositions (\ce{NdCo2}, \ce{NdNi2}, \ce{PrCo2}, \ce{PrNi2}, \ce{CeCo2}, \ce{CeNi2}), using soft x-ray absorption spectroscopy, x-ray magnetic circular dichroism (XMCD), density-functional theory, and crystal field multiplet calculations. All transition-metal moments saturate below 1~T, while the rare-earth moments do not saturate even at 5~T, consistent with van Vleck paramagnetic contributions and crystal field suppression. While the sum rules are widely used to extract element-specific magnetic moments from XMCD, we show that for 3$d$ transition metals, their application requires accurate estimates of the number of unoccupied 3$d$ states. We observe a finite magnetic moment on Ni, challenging the common assumption of its nonmagnetic character in Laves phases. The orbital magnetic moments were determined using the spin rules, while the spin moments were estimated from single-ions values from multiplet calculations, due to the invalidity of the spin sum rule for light rare-earth elements. The magnetic moments of Nd and Pr are found to be suppressed relative to their free-ion values, with multiplet theory indicating that this is due to crystal field effects. Our results confirm that Nd and Pr maintain localized $4f^3$ and $4f^2$ configurations, respectively, and that their element-specific magnetic moments are robust to rare-earth substitution. Ce, on the other hand, exhibits a tuneable mixed-valent ground state with both magnetic $4f^1$ and nonmagnetic $4f^0$ components. The relative fraction of these states varies with the electronegativity of the surrounding 3$d$ transition metals, revealing a pathway to tune Ce magnetism via composition. This work establishes a framework for accurately interpreting XMCD in light rare-earth-based intermetallics and provides insight for designing light rare-earth-based magnetocaloric materials. 

\end{abstract}

\maketitle

\section{Introduction}
\label{sec:Introduction}

Magnetocaloric hydrogen liquefaction (20-80~K) is an alternative to conventional cooling methods at cryogenic temperatures, utilizing the change in magnetic entropy upon magnetization or demagnetization of a magnetic material~\cite{franco_magnetocaloric_2018}. Intermetallics based on heavy rare-earth (HRE) elements (Gd-Lu) have been extensively studied for magnetocaloric hydrogen liquefaction due to their large magnetocaloric effect at cryogenic temperatures~\cite{romero-muniz_magnetocaloric_2023, jesla_large_2023, cwik_magnetic_2024, politova_magnetism_2024, bykov_magnetocaloric_2024}. However, these elements are expensive and classified as critical and strategically important~\cite{EU2024}, making substitution with light rare-earth (LRE) elements (La-Eu) a necessity, as they are more abundant and less costly~\cite{liu_matter_2024}. Although LREs generally exhibit weaker magnetic properties than HREs, they can still exhibit significant magnetic moments due to their localized $4f$ electrons. The electronic and magnetic properties of these individual elements can be studied using x-ray absorption spectroscopy (XAS). By varying photon energies across the $L_{2,3}$ edges of 3$d$ transition metals, $2p$ core electrons are excited into unoccupied $3d$ valence states~\cite{ohresser_magnetism_2001, van_der_laan_x-ray_2014}. Similarly, on the $M_{4,5}$ edges of rare-earth elements, $3d$ core electrons are excited into unoccupied $4f$ valence states.

X-ray magnetic circular dichroism (XMCD) can separate the orbital angular momentum $\mu_\text{L}$ and effective spin angular momentum $\mu_\text{S}^\text{eff}$. As a result of the elemental specificity of XMCD, it is possible to study the influence of substitution on the magnetic moment of the elements. A number of $5f$ actinide-, HRE-, and Sm-based cubic Laves compounds (C15, space group \ce{\textit{Fd}\Bar{3}\textit{m}}, No. 227) have previously been studied using XAS and XMCD~\cite{giorgetti_xmcd_2004, mizumaki_verification_2003, wilhelm_x-ray_2013, fujiwara_xmcd_2005, bartolome_orbital_2004, herrero-albillos_observation_2007, watanabe_pressure_2009}. In contrast, only a few studies on Nd-, Pr-, or Ce-based Laves compounds have been reported, with some exceptions such as \ce{CeFe2}~\cite{giorgetti_magnetic_1993, Delobbe_1998} and \ce{NdCo2} (only $L_{2,3}$-edges of Nd)~\cite{chaboy_relationship_2007}. Ce can occur in a mixed-valent state, fluctuating between magnetic $4f^1$ and nonmagnetic $4f^0$. The relative occupancy of these two states can be quantified using a lineshape analysis of the Ce XAS spectra~\cite{okane_magnetic_2012}.

The XMCD method uses the difference in absorption between left $\mu_+(\omega)$ and right $\mu_-(\omega)$ circularly polarized x-rays of a material saturated by an applied magnetic field~\cite{chen_experimental_1995}. $\mu_\text{L}$ and $\mu_\text{S}^\text{eff}$ can be determined by applying the sum rules to the XMCD signal~\cite{thole_3d_1985, carra_x-ray_1993}. Various studies have reported different performances of the sum rules, often due to experimental artifacts like saturation effects. Still, reliable comparisons can be made between similar compounds when measured under identical conditions in the same set-up~\cite{chen_experimental_1995, nakajima_electron-yield_1999}. Additional sources of errors can arise in systems containing multiple phases. For the $L_{2,3}$ transitions, the sum rules can be expressed as:

\begin{align}
    \mu_\text{L} &= -\langle L_z\rangle = -\left(\frac{2 q}{3 r}\right)n_{h}, \label{eq:orbitalmomentL} 
    \\
    \mu_\text{S}^\text{eff} &= 2\langle S_z\rangle+7\langle T_z\rangle = -\left(\frac{3 p - 2 q}{r}\right)n_{h}, \label{eq:spinmomentL}
\end{align}

\noindent and for the $M_{4,5}$ transitions as:

\begin{align}
    \mu_\text{L} &= -\langle L_z\rangle = -\left(\frac{q}{r}\right)n_{h}, \label{eq:orbitalmomentM} \\
    \mu_\text{S}^\text{eff} &= 2\langle S_z\rangle+6\langle T_z\rangle = -\left(\frac{5 p - 3 q}{2r}\right)n_{h}, \label{eq:spinmomentM}
\end{align}

\noindent where $p$ is the integral of the XMCD signal over the $L_3$ or $M_5$ peak, $q$ is the integrated XMCD signal over the $L_{2,3}$ or $M_{4,5}$ peaks, and $n_h$ is the number of holes in the $3d$ or $4f$ shell~\cite{chen_experimental_1995}. The $r$ is the integrated difference between the average XAS signal of $\mu_+$ and $\mu_-$ and a step function with two steps~\cite{chen_experimental_1995}. However, this approximation can lead to an underestimation of the magnetic moment, in particular for rare-earth elements~\cite{schille_experimental_1993}. The $\langle L_z\rangle$, $\langle S_z\rangle$, and $\langle T_z\rangle$ are the expectation values of the orbital, spin, and intra-atomic magnetic dipole operators, respectively. The $\langle T_z\rangle$ is generally considered negligible for powders~\cite{stohr_determination_1995}.

Determining $n_h$ is straightforward for species with integer-valued electron configurations, \textit{i.e.}, the number of 4$f$ holes of Nd and Pr. For itinerant species, however, $n_h$ cannot be measured directly. As a result, most studies omit reporting the $n_h$ used or report widely different values. For example, the number of 3$d$ holes, $n_h$, for Co in \ce{ErCo2} has been estimated as 1.1 through fitting of XMCD data to match bulk magnetization data~\cite{bartolome_orbital_2004}. In contrast, another study has obtained $n_h$ of 2.49 for elemental Co based on averaging reported theoretical calculations~\cite{chen_experimental_1995}. This resulted in magnetic moments of 0.81~$\mu_\text{B}$ and 1.75~$\mu_\text{B}$, respectively. However, the corresponding magnetic moments per hole are similar (0.70 and 0.74~$\mu_\text{B}$/$n_h$, respectively). This highlights the importance of accurately estimating $n_h$ since it linearly scales $\mu_\text{L}$ and $\mu_\text{S}^\text{eff}$ derived from the sum rules.

Using the spin sum rule, the spin moment $\mu_\text{S} = 2\langle S_z\rangle$ can be reliably determined for 3$d$ transition metals~\cite{teramura_3d, chen_experimental_1995}, while for LREs, $3d$-$4f$ exchange interactions and $3d$ core spin-split interaction induce significant deviations~\cite{teramura_4f, jo_3d4f_1997}. Furthermore, crystal field effects have a significant impact on the calculated LRE magnetic moments using the sum rules. 

We have previously investigated the magnetocaloric effect in cubic Laves phase compounds with the general formula \ce{AB2}. In that study, we focused on three different LREs, Nd, Pr, and Ce, occupying the A-site, while the B-site was occupied by an equimolar ratio of Co and Ni~\cite{lunde_machine_2025}. Magnetic measurements show that the compounds exhibit ferromagnetic ordering at low temperatures, except CeCoNi and \ce{Ce_{0.75}Pr_{0.25}CoNi}~\cite{lunde_machine_2025}. In the present work, we have used XAS and XMCD to investigate the electronic and magnetic properties of these LRE-based cubic Laves compounds in detail.

The paper is organized as follows. Section~\ref{sec:Methods} describes the experimental and computational methods used in this study. Section~\ref{sec:Results} presents the results of the structural analysis, as well as the electronic and magnetic properties of the compounds. 

\section{Methods}
\label{sec:Methods}

\subsection{Experimental}

An overview of the compounds investigated in this study is tabulated in Table~\ref{tab:samples}. All samples were prepared by arc melting and heat-treated using the protocol as previously described~\cite{lunde_machine_2025}. The compounds were stored under an Ar atmosphere to prevent oxidation. Crystal structures were investigated by x-ray diffraction (XRD) using a Bruker D2 diffractometer in Bragg-Brentano geometry (Cu-K$\alpha$ radiation, $\lambda = 1.54060$~{\AA}) at room temperature (RT). Rietveld refinements were performed using Topas V5~\cite{topas}.

\begin{table}[h!]
\centering 
\caption{\label{tab:samples} Compounds investigated in this study.}
\begin{ruledtabular}
\begin{tabular}{c c} \\ [-3 mm]
Ternary and quaternary alloys* & Binary alloys \\ [1 mm]
\hline \\ [-2.5 mm]
NdCoNi                      & \ce{NdCo_2} \\
\ce{Nd_{0.75}Pr_{0.25}CoNi} & \ce{NdNi_2} \\
\ce{Nd_{0.50}Pr_{0.50}CoNi} & \ce{PrCo_2} \\
\ce{Nd_{0.25}Pr_{0.75}CoNi} & \ce{PrNi_2} \\
PrCoNi                      & \ce{CeCo_2} \\ 
\ce{Ce_{0.25}Pr_{0.75}CoNi} & \ce{CeNi_2} \\
\end{tabular}
\end{ruledtabular}
*Samples reused from~\cite{lunde_machine_2025}.
\end{table}

XAS and XMCD measurements were carried out at the CroMag end station, DEIMOS beamline at the SOLEIL synchrotron facility, France~\cite{DEIMOS, DEIMOS2}. Data were collected in the Total Electron Yield (TEY) mode, with an energy resolution of more than $E/\Delta E=5000$, and circularly polarized light. Finely crushed sample powder was placed on an 8$\times$8~mm carbon tape attached to a Cu sample holder under ambient air. The samples were exposed to air for 5-10 minutes before being transferred to an ultra-high vacuum at a temperature of 4.2~K, and thus below the Curie temperature of the compounds~\cite{lunde_machine_2025}. The XAS spectra were recorded at $\pm$5~T. All edges (Ni, Co, Ce, Pr, Nd) were measured for all samples and binaries listed in Table~\ref{tab:samples}, except Ni, which was measured only for NdCoNi, PrCoNi, and the binaries. Typically, 18 scans were performed per set of edges, and their average was determined, normalized by their pre-edge. Finally, the XAS and XMCD spectra were calculated from the average and difference between $\mu_+$ and $\mu_-$, respectively.

For each polarization, field-dependent measurements were performed at 4.2~K from $-$5~T to $+$5~T at the energy corresponding to the strongest XMCD signal of the $L_{3}$ or $M_{5}$ edges. The energies used for Co, Ni, Ce, Pr, and Nd were 781.8, 855.9, 903.3, 953.4, and 1004.1~eV, respectively. The corresponding magnetization curves were constructed from the field-dependent XMCD signal and scaled using the total magnetic moments obtained from the sum rules at 5~T. The data points close to zero field were removed since they resulted from strong perturbations of TEY in this range.

\subsection{Computational}

\emph{Ab initio} spin-polarized density-functional theory (DFT) calculations were performed with the Vienna \emph{Ab Initio} Simulation Package, \textsc{vasp} 6.4.1~\cite{kresse_efficiency_1996, kresse_ab_1993, kresse_efficient_1996}, within the scalar relativistic approximation including spin-orbit coupling using the SCAN-L~\cite{SCAN-L} meta-GGA functional. The Brillouin zone was sampled using a Monkhorst-Pack~\cite{monkhorst_special_1976} $11 \times 11 \times 11$ $k$-point grid, and a plane-wave cutoff energy of 550~eV was employed. A total of 60 bands were included in the calculation, with electronic self-consistency achieved to a convergence threshold of $10^{-6}$~eV per supercell. Structural optimization with a force tolerance criterion of $0.01$~eV/Å was conducted for all the materials. To correct for on-site electron–electron interactions in localized $d$- and $f$-orbitals, the Hubbard U method was employed. The values of U were estimated using the linear response approach of Cococcioni and de Gironcoli~\cite{cococcioni_linear_2005}.  

Full multiplet calculations within the crystal field regime were constructed using \textsc{quanty}~\cite{haverkort_bands_2014}. Values for the Slater integrals and spin-orbit coupling constants were obtained from Cowan's Hartree-Fock code~\cite{cowan_theoretical_1968} and subsequently scaled by 90\% of their values to account for configuration interaction effects. The coefficients were optimized to reproduce the XMCD spectra and were restricted by the orbital moment calculated from the orbital sum rules. For a more detailed description of the methodology, refer to the Supplemental Material~\cite{supplemental}.

\section{Results} 
\label{sec:Results}

\subsection{Structural analysis}
\label{sec:Structural}

The \ce{Nd_{1-x}Pr_xCoNi} (0~$\leq$~x~$\leq$~1) and \ce{Ce_{0.25}Pr_{0.75}CoNi} compounds investigated in this work were reused from~\cite{lunde_machine_2025}. Combined XRD and SEM-EDS analyses confirmed that the compounds exhibit the cubic Laves phase with minor \ce{AB} or \ce{AB3} secondary phases. The XRD results of the binaries synthesized for this work are reported in the Supplemental Material~\cite{supplemental}. Rietveld refinements confirm that all compounds primarily consist of the cubic \ce{AB2} Laves phase. Nd- and Pr-based binaries exhibit some secondary \ce{AB} or \ce{AB3} phases (0.8-6.5~wt\%). The Ce-based compounds exhibit significant oxidation due to air exposure, consistent with previous observations for CeCoNi~\cite{lunde_machine_2025}.

\subsection{Electronic properties}
\label{sec:Electronic}

\begin{figure*}[t] 
  \centering
  \begin{subfigure}[b]{0.48\textwidth}
    \includegraphics[width=\textwidth]{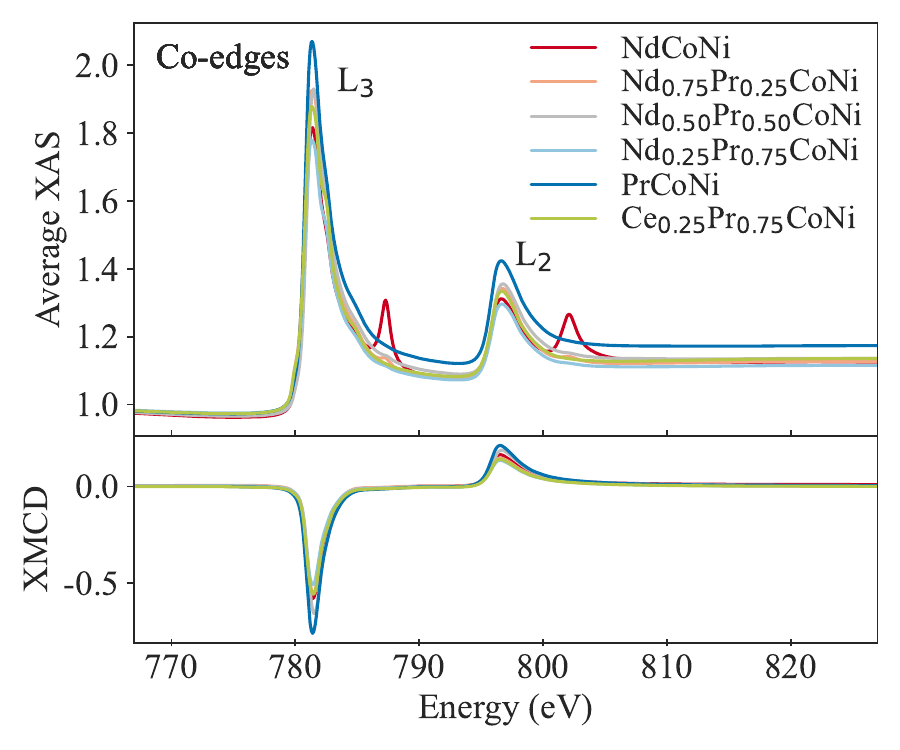}
    \caption{}
    \label{fig:XMCD_Co}
  \end{subfigure}
  \begin{subfigure}[b]{0.48\textwidth}
    \includegraphics[width=\textwidth]{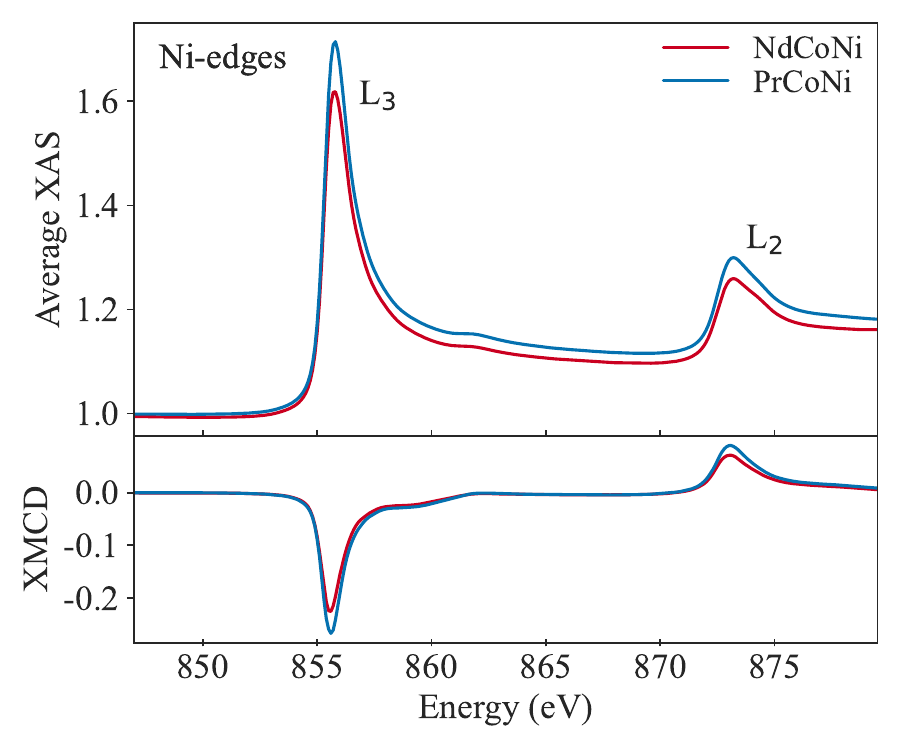}
    \caption{}
    \label{fig:XMCD_Ni}
  \end{subfigure}
  \\
  \begin{subfigure}[b]{0.48\textwidth}
    \includegraphics[width=\textwidth]{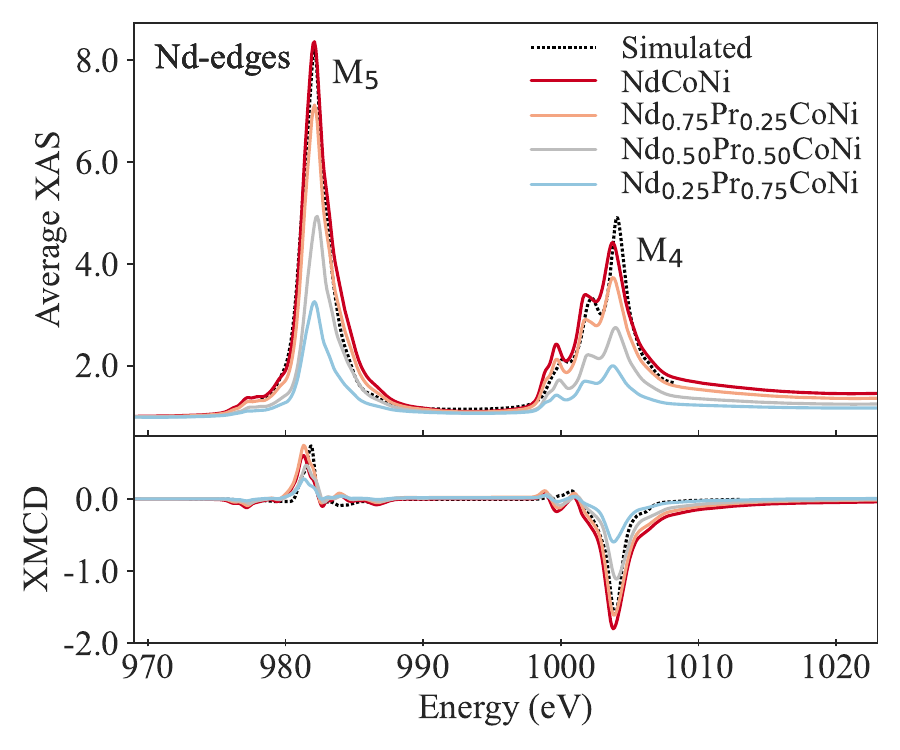}
    \caption{}
    \label{fig:XMCD_Nd}
  \end{subfigure}
  \begin{subfigure}[b]{0.48\textwidth}
    \includegraphics[width=\textwidth]{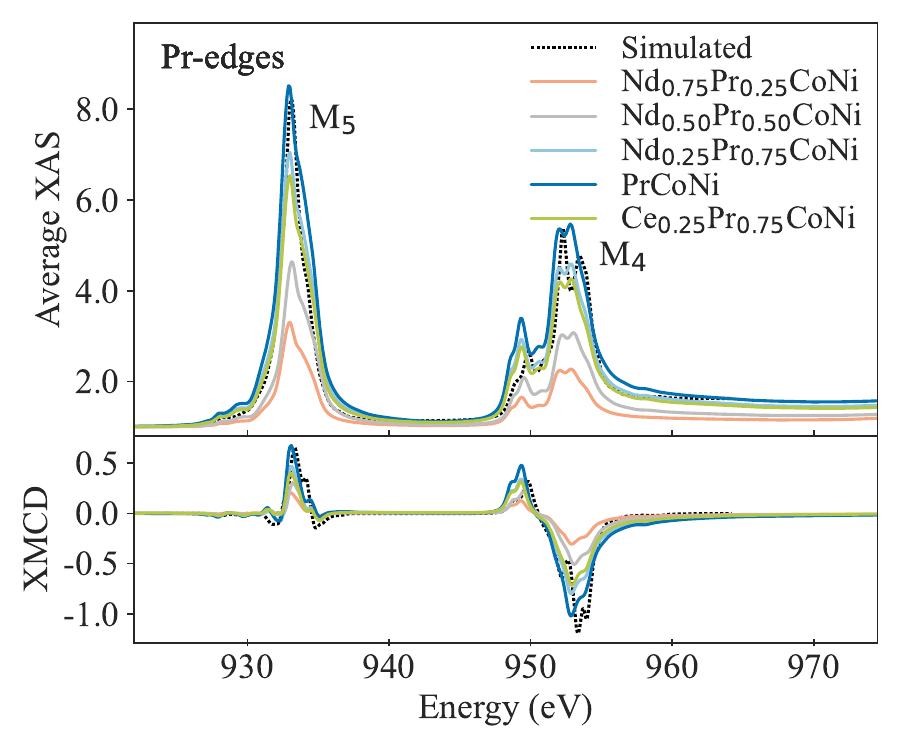}
    \caption{}
    \label{fig:XMCD_Pr}
  \end{subfigure}
  \caption{(a) Co and (b) Ni experimental $L_{2,3}$ and (c) Nd and (d) Pr simulated and experimental $M_{4,5}$ edges measured at 4.2~K and 5~T. The XAS spectra represents the average of the left and right polarized x-ray absorption, while XMCD shows the difference.}
  \label{fig:XMCD}
\end{figure*}

The XAS and corresponding XMCD spectra for the $L_{2,3}$ edges of Co and Ni and the $M_{4,5}$ edges of Nd and Pr are presented in Fig.~\ref{fig:XMCD} and in the Supplemental Material~\cite{supplemental}. The Co-edges (Fig.~\ref{fig:XMCD}(a)) display $L_3$ and $L_2$ peaks at 781.3~eV and 796.5~eV, respectively. As expected, the $L_3$ peak is more intense. A small shoulder can be observed on the right side of the $L_3$ peak, possibly as a result of surface oxidation. Two additional peaks appear at energies higher than the main peaks of Co (787.3~eV and 802.1~eV) in the NdCoNi spectrum, and weaker peaks at the same energies are barely visible in the \ce{Nd_{0.75}Pr_{0.25}CoNi} spectrum. These features appear at the energies corresponding to Ba with the expected branching ratio, making Ba contamination from the sample holder a possible explanation. As these features are assumed to result from contamination, their contributions to the XAS signal were not included in the sum rule analysis and could not influence the results. Aside from these additional features, the Co XAS spectra are consistent across all samples.

The Ni-edges of NdCoNi and PrCoNi (Fig.~\ref{fig:XMCD}(b)) show the $L_3$ and $L_2$ peaks at 855.8 and 873.2~eV, respectively. The XAS signal does not differ between these two samples or their corresponding binary samples, indicating no significant variation in the Ni electronic environment. A minor satellite peak is present at a higher energy than the $L_3$ peak ($\sim$862~eV) for all samples. This was also observed for Ni in the cubic Laves phase \ce{GdNi2}~\cite{mizumaki_verification_2003}, which was attributed to the $2p^5d^9$ configuration.

The $M_{4,5}$ edges of Nd and Pr (Fig.~\ref{fig:XMCD}(c-d)) exhibit multiple peaks, which is common for $4f$ elements due to the strong $4f$ localization giving rise to 3$d$-4$f$ and $4f$-$4f$ interaction, resulting in a multiplet-level splitting of the quasi-atomic initial and final states~\cite{thole_3d_1985}. The multiplet structure appears systematic for the series. The multiplets are well resolved, indicating a lack of $4f$-$3d$ hybridization. This is further confirmed by DFT calculations, which indicate that the $f$-electron valency is given as an integer number for both Nd and Pr. Furthermore, the partial density of states plot for \ce{Nd_{0.50}Pr_{0.50}CoNi}, found in the Supplementary Material~\cite{supplemental}, reveals highly localized $f$-electrons with minimal overlap with the Co and Ni $d$-states.

For all the Nd-edges (Fig.~\ref{fig:XMCD}(c)), the $M_5$ peak at 982.1~eV features a small low-energy shoulder, while the $M_4$ edge displays multiple peaks at 999.7, 1001.8, and 1003.8~eV. The XAS spectra are consistent for all samples, including the binaries. Nd exhibits a $4f^3$ electron configuration, according to reported calculations and experiments with identical XAS spectra~\cite{thole_3d_1985, goedkoop_calculations_1988, laan_identification_1986}. Furthermore, the Pr-edges (Fig.~\ref{fig:XMCD}(d)) show an $M_5$ peak (932.9~eV) with a shoulder at lower energies and an $M_4$ edge consisting of multiple peaks (949.3, 952.0, and 952.8~eV). Again, the signal is similar for all samples. Reported calculations and experiments indicate that Pr exhibits a $4f^2$ electron configuration~\cite{thole_3d_1985, goedkoop_calculations_1988, laan_identification_1986}. The electron configurations and spectra of Nd and Pr were confirmed using DFT calculations and multiplet theory simulations of the XAS and XMCD spectra.

\subsection{Magnetic properties}
\label{sec:Magnetic}

All measured edges (Co, Ni, Ce, Pr, Nd) for the studied compounds produce XMCD signals, as seen in Fig.~\ref{fig:XMCD}. The strongest XMCD peaks relative to the strength of the XAS were observed at the Co-edge, followed by the Nd-, the Pr-, Ni-, and Ce-edges, respectively.

\begin{figure*}[t]
  \centering
  \subfloat[][]{\includegraphics[width = 0.5\textwidth]{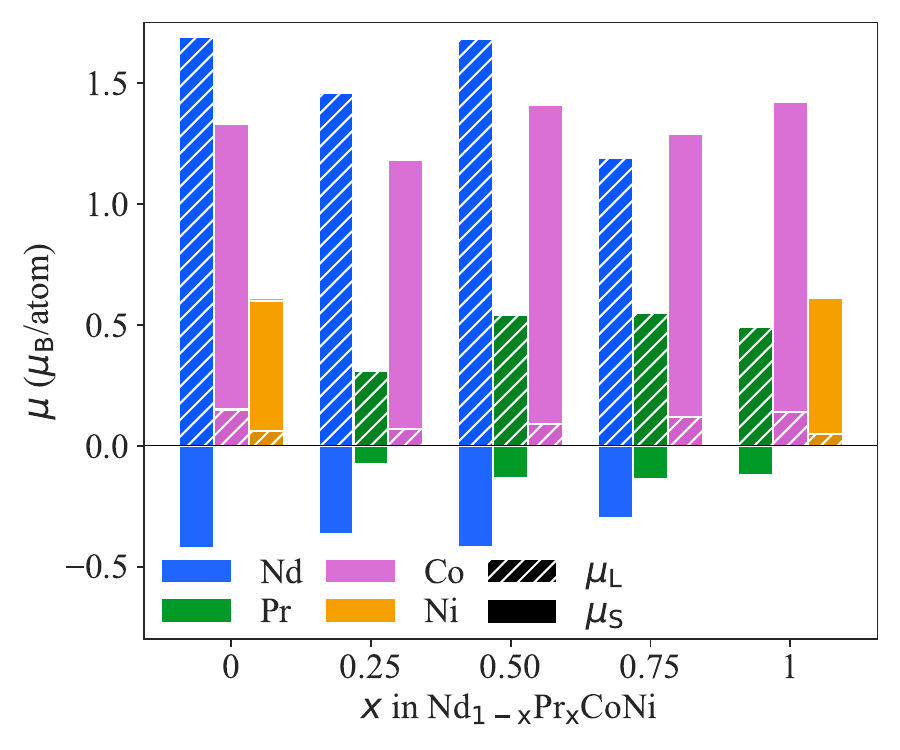} \label{fig:Sum_rules_all}}   
  \subfloat[][]{\includegraphics[width = 0.5\textwidth]{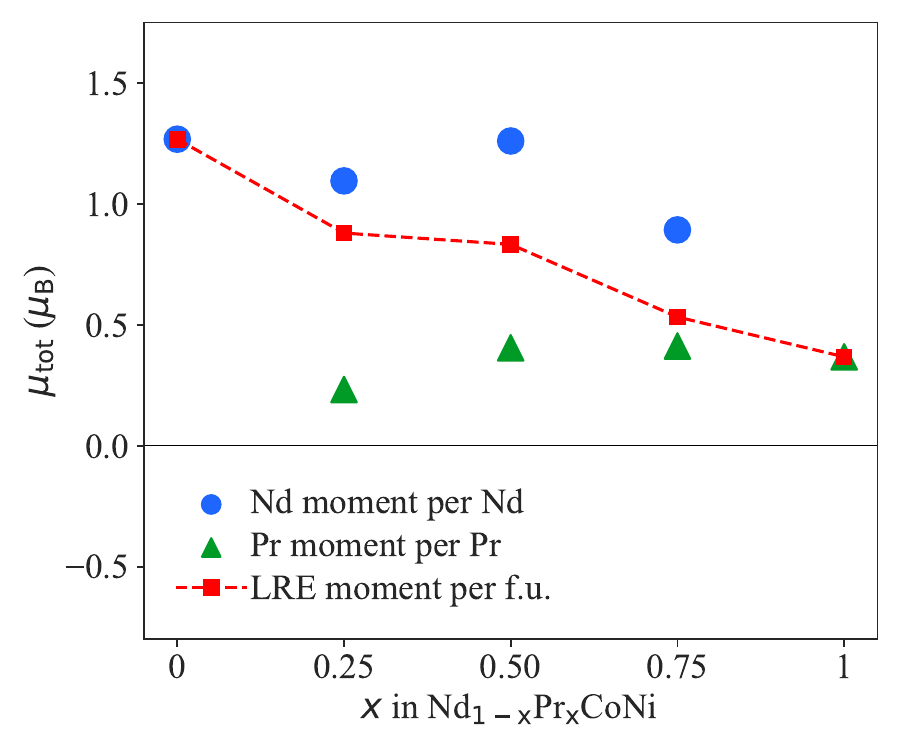} \label{fig:Sum_rules_LRE}} \\
  \caption{(a) The orbital (hatched) and spin (filled) moment for Nd, Pr, Co, and Ni. Ni was only measured for NdCoNi and PrCoNi. $\mu_\text{S}$ for Pr and Nd was calculated using multiplet theory, while all other values are found by applying the sum rules to the experimental XMCD data. (b) The total magnetic moment of Nd and Pr per atom, together with the total LRE moment per formula unit.} \label{fig:Sum_rules}
\end{figure*}

\begin{figure*}[t]
  \centering
  \subfloat[][]{\includegraphics[width = 0.49\textwidth]{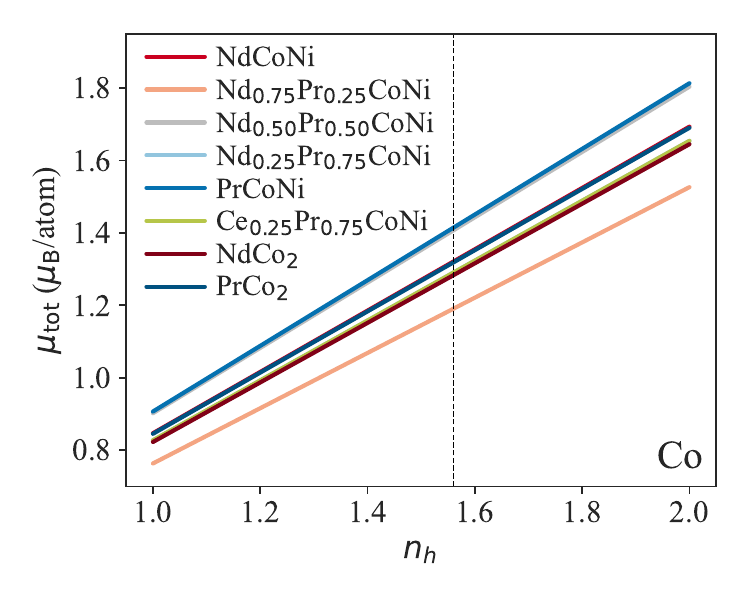} \label{fig:n_Co}} 
  \subfloat[][]{\includegraphics[width = 0.49\textwidth]{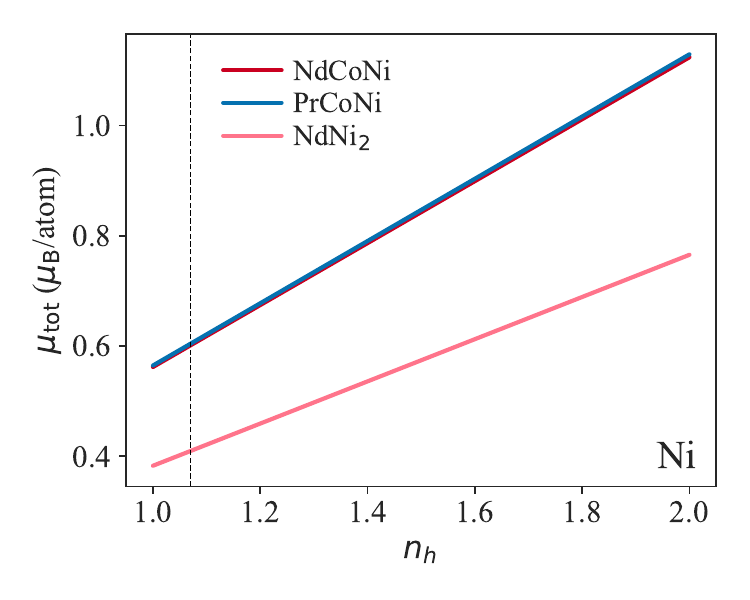} \label{fig:n_Ni}} \\
  \caption{The total Co and Ni moment as a function of the $n_h$ chosen in the sum rules. The $n_h$ determined using DFT calculations in this work are shown with a vertical line for each of the edges.} \label{fig:holes}
\end{figure*}

Figure~\ref{fig:Sum_rules} shows the results of the sum rule calculations for \ce{Nd_{1-x}Pr_xCoNi}, while the magnetic moments for all studied compounds are tabulated in Supplemental Material~\cite{supplemental}. $\langle T_z\rangle=0$ is assumed for the sum rule analysis since the compounds are powders~\cite{stohr_determination_1995}. The Pr and Nd $\mu_\text{S}$ were estimated from the single-ion values from our crystal electric field model due to the invalidity of the spin sum rule for LREs, and values obtained using this sum rule are tabulated in the Supplemental Material~\cite{supplemental} for comparison. The model assumes a cubic crystal field and was constrained by the XMCD data and the orbital moment calculated according to the orbital sum rule. All other values are found by applying the sum rules to experimental XMCD data. The magnetic moments were only calculated for the binary samples expected to exhibit magnetic ordering at 4.2~K, i.e., \ce{NdCo2}, \ce{NdNi2}, and \ce{PrCo2}, and not for the paramagnetic compounds (\ce{PrNi2}, \ce{CeCo2}, and \ce{CeNi2})~\cite{farrell_magnetic_1966}. Figure~\ref{fig:Sum_rules}(a) shows $\mu_\text{L}$ and $\mu_\text{S}$ for Nd, Pr, Co, and Ni in \ce{Nd_{1-x}Pr_xCoNi}. Figure~\ref{fig:Sum_rules}(b) shows the total magnetic moments of Pr and Nd for \ce{Nd_{1-x}Pr_xCoNi} together with the summed LRE magnetic moments per formula unit. All elements exhibit a finite magnetic moment. There is no clear trend for the Co magnetic moment based on the LRE composition, and the variation between the samples of $\pm$0.10~$\mu_\text{B}$/atom is less than the error bar associated with XMCD and sum rule analysis. The magnetic moment of Ni is similar for NdCoNi and PrCoNi (0.60 and 0.61~$\mu_\text{B}$, respectively).

The $n_h$ for 3$d$ transition metal itinerant species is typically obtained by subtracting the number of $d$-electrons from the total number of available $d$-states. We calculated the number of electrons from the partial electronic density of states calculated by DFT. When subtracting these values from the number of $d$-states, we obtain values of 2.66 for Co and 1.34 for Ni. These values are similar to some values reported in the literature, \emph{e.g.} 2.49 and 1.59 for Co and Ni, respectively ~\cite{chen_experimental_1995, mizumaki_verification_2003}. However, this approach overestimates the magnetic moment for our samples compared to magnetization data.

If we calculate $n_h$ by projecting the DFT ground-state electron density onto spherical harmonics and integrating the partial electronic density of states above the Fermi level, we obtain lower values, around 1.56 for Co and 1.07 for Ni. The latter approach is physically more meaningful as it is $n_h$ we aim to obtain. It is thus more correct, and we use these values in our analysis. The dependence of the moment as a function of $n_h$ for these elements is shown in Fig.~\ref{fig:holes}, highlighting the importance of an accurate estimation of this parameter for comparing moments across different studies and experimental techniques.

The $n_h$ for Pr and Nd were determined to be integer numbers of 12 and 11, respectively, based on the XAS spectra and DFT calculations.

\begin{figure*}[htbp]
    \centering
    \includegraphics[width = 0.97\textwidth]{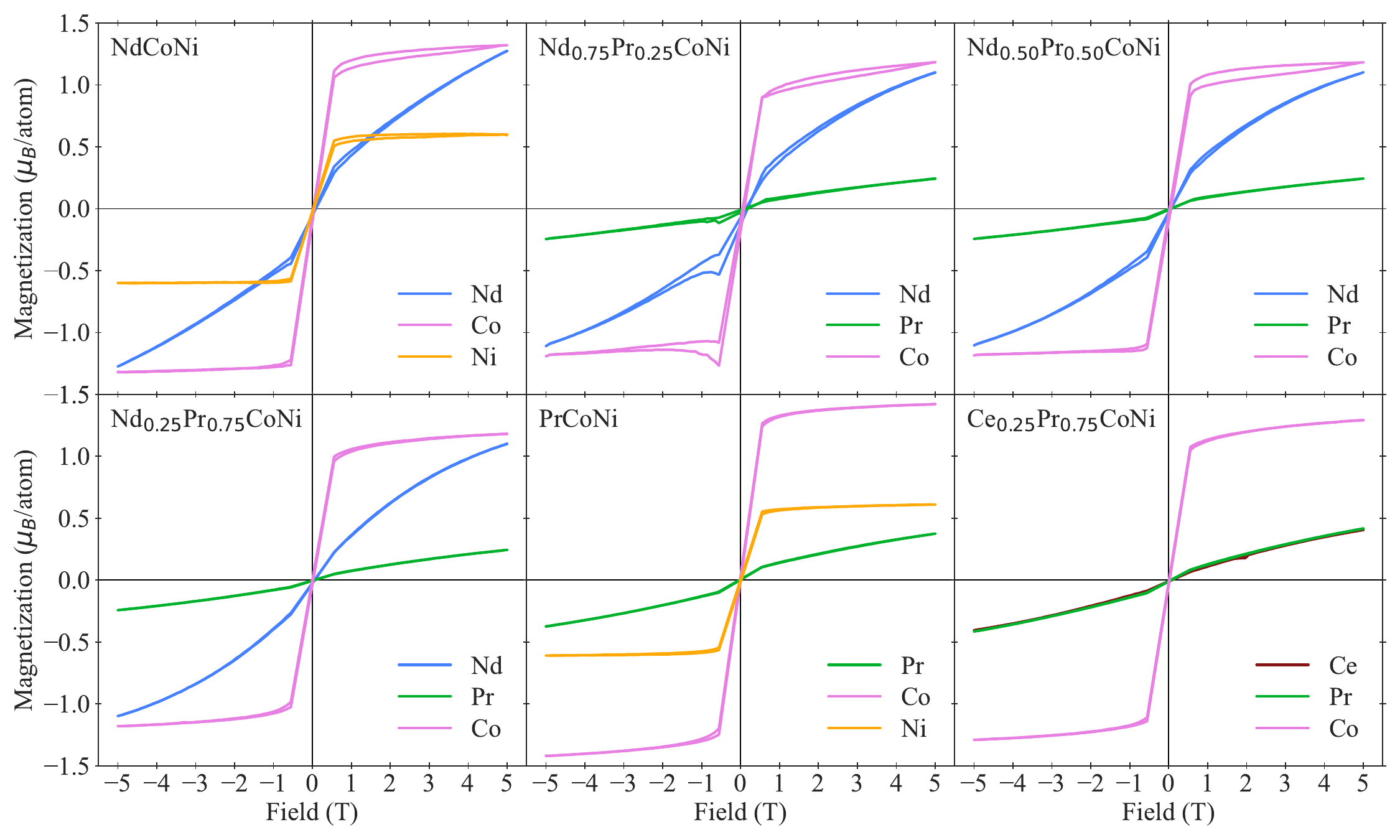}
    \caption{Field-dependent XMCD measurements performed at 4.2~K, at the energy of the strongest peak for each edge. The curves are scaled using the results from the sum rules at 5~T. Data points close to zero field have been removed.}
    \label{fig:Hysterese}
    
    \vspace{0cm} 

    \centering
    \includegraphics[width = 0.97\textwidth]{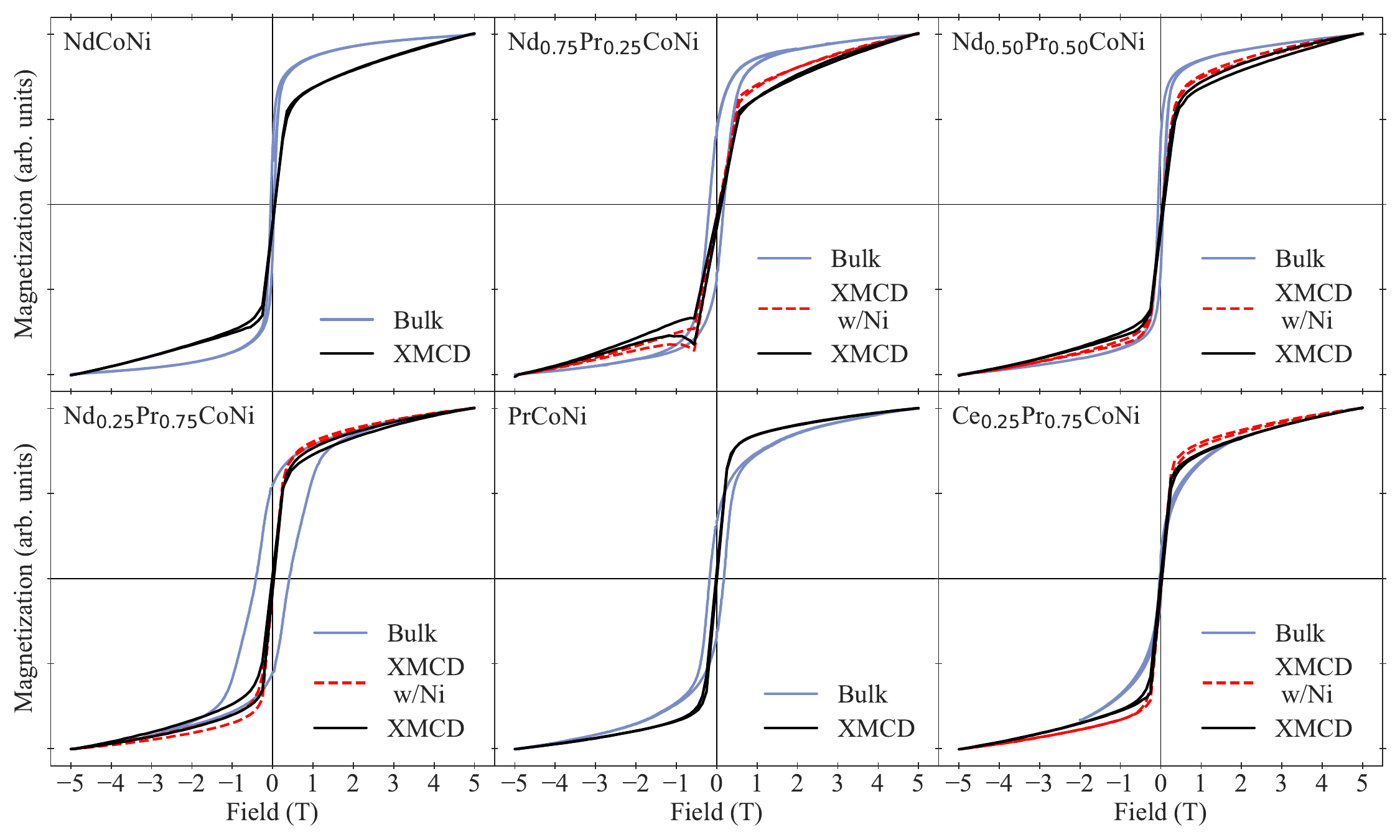}
    \caption{Normalized field-dependent magnetization curves measured using bulk magnetization measurements at 5~K~\cite{lunde_machine_2025} and XMCD, as the sum of the field-dependent magnetization of all the measured edges in Fig.~\ref{fig:Hysterese}. For compounds where the Ni edge was not measured, the Ni contribution from NdCoNi is added as dashed curves.}
    \label{fig:Hysterese_PPMS}
\end{figure*}

The magnetic moments of Co and Ni saturate around 0.5~T, while those of the LREs remain unsaturated at 5~T, the maximum field applied, as seen in Fig.~\ref{fig:Hysterese}. Consequently, the results derived using the sum rules in this work correspond to the values at 5~T, which do not represent the full saturation moments for the LREs. TEY was used as a probe, measuring both the 3$d$ transition metal and LRE edges on the order of a few nanometers, making the results sensitive to the surface, regardless of the element. The lack of saturation in the magnetization may be attributed to van Vleck paramagnetism, which arises from the mixing of higher multiplets with the ground state, yielding a linear contribution as a function of the applied field~\cite{wallace1968magnetic}.

The total magnetization of the elements in the compounds is compared to bulk magnetization measurements from our previous study~\cite{lunde_machine_2025} in Fig.~\ref{fig:Hysterese_PPMS}. Ni saturates at low field (Fig.~\ref{fig:Hysterese}), and when this contribution is added to the samples where the Ni edge was not measured, the XMCD signal for the samples becomes slightly more saturated.
The remaining discrepancies between the two measurement techniques may be attributed to mild surface oxidation at ambient temperature during sample handling, forming a poorly crystalline layer of antiferromagnetic or paramagnetic contributions. However, the absence of an exchange bias shift in the XMCD curves suggests oxidation to be limited~\cite{avila-gutierrez_influence_2023}.

There is a difference in hysteresis between bulk and XMCD magnetization for some of the samples. Part of this discrepancy can be attributed to that data points close to zero field have been removed from the XMCD results due to strong perturbations of TEY at these values. However, for \ce{Nd_{0.75}Pr_{0.25}CoNi}, the difference is larger than what can be attributed to this explanation. This difference could be caused by the differences between the two measurement techniques, involving probing either the bulk or the surface. The exact same samples were characterized using the two techniques, eliminating the possibility of differences in saturation and hysteresis due to poor reproducibility in the sample preparation.

The magnetic moments of Co and Ni (1.2-1.4 and 0.4-0.6~$\mu_\text{B}$, respectively) are significantly smaller than those calculated from the free ions using Hund's rules (6.75 and 6.00~$\mu_\text{B}$, respectively). The magnetism of the $3d$ transition metals results from unpaired $3d$-electrons, and $\mu_\text{L}$ is often quenched compared to what is calculated for the magnetic ground state using Hund's rules~\cite{blundell_magnetism_2001}. This was observed for Co and Ni, where $\mu_\text{L}$ is close to zero, and the moment mainly results from $\mu_\text{S}$. Transition metal magnetic moments generally couple ferromagnetically to the LRE moments in cubic Laves compounds but antiferromagnetically to HRE moments~\cite{politova_magnetism_2024, farrell_magnetic_1966, ermolenko_compositional_2019}. Ferromagnetic coupling is observed in this system, for all elements exhibiting $\mu_\text{tot}$ parallel to the applied field. The LRE moment is not constant upon transition metal substitution.

As expected, Co exhibits a higher magnetic moment than Ni. However, Ni is generally assumed to exhibit a small or no magnetic moment in cubic Laves compounds~\cite{bykov_magnetocaloric_2024}, and has been reported as nonmagnetic for \ce{CeNi2}, \ce{NdNi2}, and \ce{PrNi2}~\cite{farrell_magnetic_1966}. In this study, we observe a finite magnetic moment for Ni in all evaluated compounds, with a $\mu_\text{tot}$ of 0.41~$\mu_\text{B}$ for \ce{NdNi2}, contradictory to previous reports~\cite{farrell_magnetic_1966}. For \ce{GdNi2} Laves phases, Ni has a moment of 0.2~$\mu_\text{B}$ antiparallel to Gd, determined using XMCD at 1.4~T and 25~K~\cite{mizumaki_verification_2003}. This magnetic moment is significantly smaller than what we have observed (0.4-0.6~$\mu_\text{B}$), but this can be attributed to the differences in the applied field and temperature being used. To the best of our knowledge, the magnetic moment of Co in LRE Laves phases has not previously been determined using the sum rules. However, for \ce{ErCo2} a Co moment of 0.9-1.0~$\mu_\text{B}$ has been found at 1~T~\cite{herrero-albillos_observation_2007, bartolome_orbital_2004}. The Co magnetic moment of \ce{NdCo2} has been studied using neutron diffraction, where the Co moment has been determined in the range 0.51-0.86~$\mu_\text{B}$~\cite{xiao_canted_2006}. As previously discussed, part of these discrepancies could be attributed to different choices of $n_h$.

For the LREs, $\mu_\text{L}$ is significantly higher than $\mu_\text{S}$ and oppositely directed, as expected based on Hund's rules. Magnetic moments calculated using Hund's rules for free trivalent ions, $\mu=g_JJ\mu_\text{B}$, for Nd (3.27~$\mu_\text{B}$) and Pr (3.20~$\mu_\text{B}$) are higher than the magnetic moments derived in this work, 0.89-1.26~$\mu_\text{B}$ and 0.24-0.41~$\mu_\text{B}$, respectively. The low values for the LREs are most likely due to the lack of saturation in the magnetization at 5~T (see Fig.~\ref{fig:Hysterese}). A further reduction of the Pr moment compared to the Nd magnetic moment can be attributed to crystal field effects. Nd has a significantly higher $\mu_\text{tot}$ than Pr, resulting in an almost linear decrease in $\mu_\text{tot}$ per formula unit for \ce{Nd_{1-x}Pr_xCoNi} (0~$\leq$~x~$\leq$~1) with increasing Pr substitution (Fig.~\ref{fig:Sum_rules}(b)). The magnetic moment per Nd and Pr atom is quite similar for all the compounds and is not significantly influenced by the substitution. For comparison, the $\mu_\text{S}$ values of Pr and Nd obtained using the invalid LRE spin sum rule are significantly higher than those obtained using multiplet theory calculations. This results in a negative $\mu_\text{tot}$ for all compounds using these values.

Bulk magnetization data for PrCoNi and NdCoNi yield magnetic moments of 2.2 and 2.4~$\mu_\text{B}$, respectively. In comparison, the total moments obtained from the XMCD data for these compounds are 2.5 and 3.6~$\mu_\text{B}$, respectively. Consequently, the moments derived from XMCD data are overestimated compared to the bulk magnetization data. The magnetic moments obtained from XMCD might be influenced by surface oxidation due to the low penetration depth of TEY. Possible surface oxides of Pr, Nd, Co, and Ni are paramagnetic or antiferromagnetic at the measurement temperature of 4.2~K~\cite{rai_magnetism_2020,hinatsu_magnetic_1988, roth_magnetic_1964,mandziak_tuning_2019}. However, the larger magnetic moments obtained by XMCD compared to bulk magnetization could indicate that any surface oxidation is likely limited.

\subsection{Cerium electronic and magnetic properties}
\label{sec:Cerium}

\begin{figure}[t]
  \centering
  {\includegraphics[width = 0.49\textwidth]{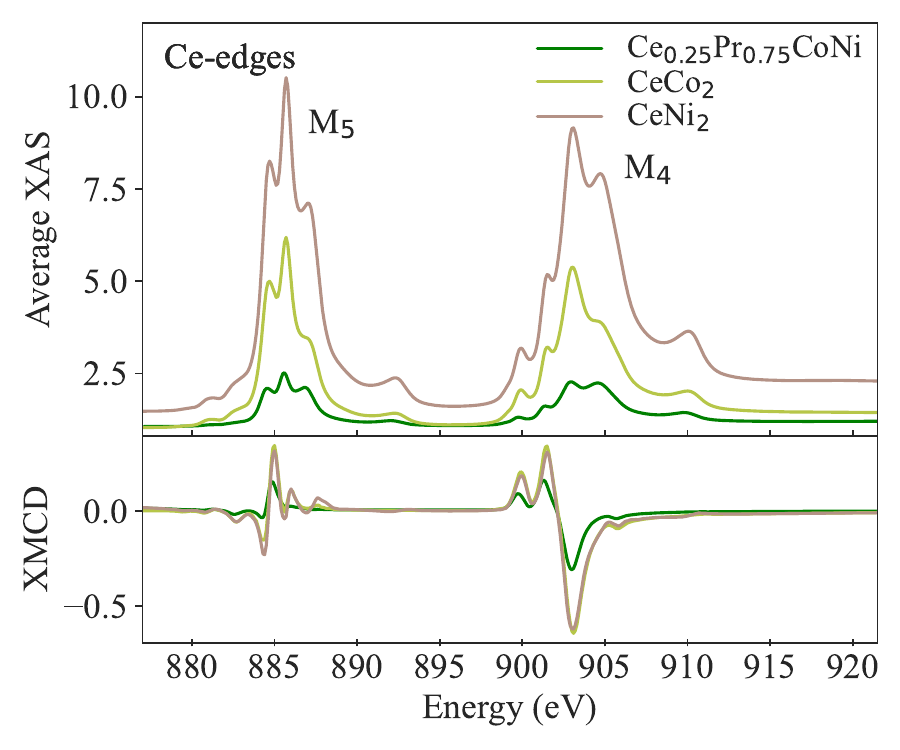} \label{fig:XMCD_Ce}} \\
  \caption{Ce $M_{4,5}$ edges measured at 4.2~K and 5~T. XAS shows the average of the right and left polarized x-ray absorption, while XMCD shows the difference.} \label{fig:Ce-edge}
\end{figure}

Figure~\ref{fig:Ce-edge} presents the XAS and XMCD spectra of the Ce-edge for \ce{Ce_{0.25}Pr_{0.75}CoNi}, with the binary samples \ce{CeCo2} and \ce{CeNi2} included for comparison. Both the $M_5$ and $M_4$ edges consist of multiple peaks. The three most intense $M_5$ peaks are at 884.7, 885.7, and 887.1~eV, and the two most dominant $M_4$ peaks are at 903.0 and 904.7~eV. Both edges contain shoulders at higher and lower energies than the main peaks. 

Some of the peaks correspond to magnetic $4f^1$ ($J=\sfrac{5}{2}$) Ce states, while others correspond to nonmagnetic $4f^0$ ($J=0$) Ce states, resulting in changing XMCD signals. The XAS signal evolves significantly along the series (\ce{Ce_{0.25}Pr_{0.75}CoNi}, \ce{CeCo2}, and \ce{CeNi2}) as the ratio between the different peaks varies, indicating changes in the $4f^1$/$4f^0$ ratio. The $4f^0$ are shifted to higher energies for all compounds to reduce electron screening, resulting in tighter binding to the core levels~\cite{melcher_cerium_2005}. 

\begin{figure*}[t]
    \centering
    \includegraphics[width = \textwidth]{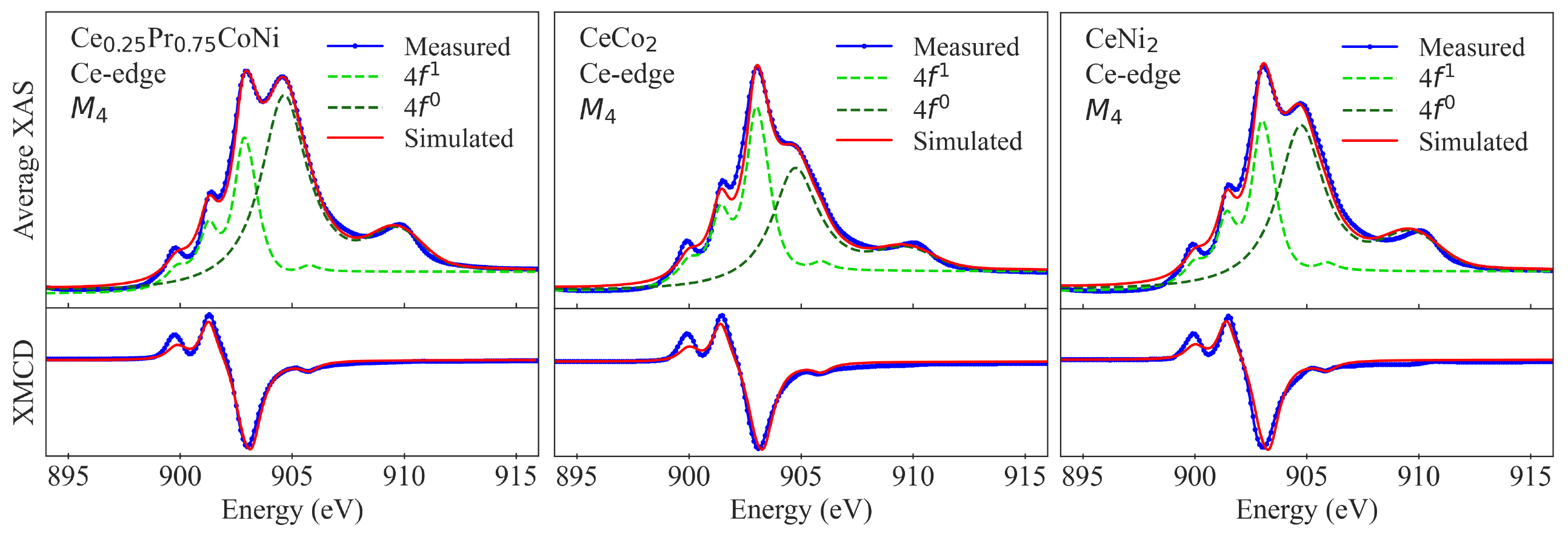}
    \caption{Lineshape analysis of the experimental $M_4$ edge of Ce, used to determine the fraction of $4f^0$ and $4f^1$ using multiplet theory simulations. The simulated curve is the sum of the $4f^0$ and $4f^1$ contributions, fitted to the background-corrected experimental data. For the XMCD signal, only $4f^1$ contributes to the signal and is simulated.}
    \label{fig:Lineshape}
\end{figure*}

To approximate the $4f^1$/$4f^0$ ground state ratio, a lineshape analysis was employed on the $M_{4}$ edges~\cite{vasili_direct_2017}, shown in Fig.~\ref{fig:Lineshape}. It should also be underlined that the lineshape analysis does not differentiate between Ce $4f^0$ and $4f^1$ in the main Laves phases and in a secondary phase. The average XAS and the XMCD spectra were fitted using spectra simulated from the Ce $4f^1$ and Ce $4f^0$ configurations, as calculated through multiplet theory. These spectra were integrated to determine the fractions of $4f^1$ and $4f^0$. This analysis revealed a $4f^1$:$4f^0$ ratio of 1:1 for \ce{CeCo2}, 2:3 for \ce{CeNi2}, and 1:2 for \ce{Ce_{0.25}Pr_{0.75}CoNi}. This confirms the high presence of nonmagnetic Ce $4f^0$ in the Ce-containing samples, as expected based on the small lattice constants and low bulk magnetization of \ce{Ce_{1-x}Pr_xCoNi} (0.25~$\leq$~x~$\leq$~1) compounds in our previous study~\cite{lunde_machine_2025}. This is further supported by the smaller lattice constants of \ce{CeCo2} (7.16~\AA) and \ce{CeNi2} (7.22~\AA) compared to \ce{PrCo2} (7.31~\AA) and \ce{PrNi2} (7.29~\AA) in the present study. Liu \textit{et al.}~\cite{liu_designing_2023} observed increasing lattice parameters from Pr to Ce in \ce{Pr_{1-x}Ce_xAl_2}, indicating a dominant $4f^1$ configuration. This suggests Ce exhibits a higher fraction of $4f^1$ configuration in cubic Laves compounds with Al at the B-site than with Ni and Co at the B-site. This could be a result of the higher electronegativity of Co (1.88) and Ni (1.91) compared to Al (1.61). The Ce-edge of \ce{Ce_{0.25}Pr_{0.75}CoNi} exhibits the strongest XMCD signal relative to the XAS signal out of the three compounds, despite the lowest fraction of $4f^1$ configuration. This is possibly due to the fact that this compound is ferromagnetically ordered. Both \ce{CeCo_{2}} and \ce{CeCo_{2}} are paramagnetic at the measured temperature (4.2~K), while \ce{Ce_{0.25}Pr_{0.75}CoNi} orders ferromagnetically at 11~K, which will give a stronger ordering of the magnetic moments and provide a higher XMCD signal.

For the sum rules, the $4f^1$ configuration is assumed for Ce, since the $4f^0$ configuration does not contribute to the magnetic moment, resulting in $n_h=13$. This choice of $n_h$ is consistent with what has been published for mixed-valent Ce~\cite{vasili_direct_2017}. As previously pointed out, the spin sum rule is not reliable for LREs. Additionally, the spin sum rule (Eq~\ref{eq:spinmomentM}) requires well-separated $M_{4}$ and $M_{5}$ edges, which is not the case for Ce (Fig.~\ref{fig:Ce-edge}(a)). Therefore, only $\mu_\text{L}$ was determined experimentally using the sum rules. As for Nd and Pr, the $\mu_\text{L}$ (0.53~$\mu_\text{B}$) is parallel to the applied field, and the calculated $\mu_\text{S}$ (0.13~$\mu_\text{B}$) is antiparallel and smaller, resulting in a $\mu_\text{tot}$ (0.40~$\mu_\text{B}$) parallel to the field. The Ce moment is calculated per Ce, not per Ce $4f^1$. There is a surprisingly small difference in magnetic moment between Pr and Ce, considering the difference measured in the compounds using magnetization measurements~\cite{lunde_machine_2025}. Ce $4f^1$ has a calculated $\mu$ using Hund's rules of 2.14~$\mu_\text{B}$, while Ce $4f^0$ has 0~$\mu_\text{B}$~\cite{blundell_magnetism_2001}. Based on the fraction of Ce $4f^1$ and Ce $4f^0$ estimated from the lineshape analysis, the low magnetic moment determined from the sum rules indicates that the Ce moment is suppressed by the lack of saturation and crystal field effects, similarly to the Pr and Nd moments. 

\section{Conclusions}

In this work, we used soft x-ray absorption spectroscopy to investigate the electronic and magnetic properties of LRE-based \ce{Nd_{1-x}Pr_xCoNi} (0~$\leq$~x~$\leq$~1) and \ce{Ce_{0.25}Pr_{0.75}CoNi} cubic Laves phase compounds together with the corresponding binary compounds for magnetocaloric applications. The experimental results are supported by density-functional theory calculations, which enable the calculation of the number of holes.

Ni and Co exhibited a slightly larger magnetic moment than expected (0.4-0.6 and 1.2-1.4~$\mu_\text{B}$, respectively), whereas the magnetic moments of the LRE elements were suppressed by the crystal field. The magnetic moments of the 3$d$ transition metals saturate at low magnetic fields. In contrast, the magnetic moments of the LRE elements did not reach a flat saturation plateau at 5~T. This can be attributed to van Vleck paramagnetism, which provides a linear contribution as a function of the field.

We found that Ce in the cubic Laves compounds consists of between 50 and 66\% of nonmagnetic Ce $4f^0$ ($J=0$) and between 50 and 34\% of magnetic Ce $4f^1$ ($J=\sfrac{5}{2}$), with the fractions being dependent on the electronegativity of the 3$d$ transition metals in the compound. This tunability of magnetic Ce is important as Ce is the most abundant of the LRE elements. 

\begin{acknowledgments}
This work was financed by the Research Council of Norway under project number 336403. Experiments were performed on the DEIMOS beamline at SOLEIL Synchrotron, France (proposal number 20240548). We are grateful to the SOLEIL staff for their smooth operation of the facility. Calculations were performed on resources provided by Sigma2---the National Infrastructure for High Performance Computing and Data Storage in Norway, Grant No.~NN11060K. 
\end{acknowledgments}

\section*{Data Availability}

The data that support the findings of this article are openly available~\cite{PT4R3F_2025}.

\bibliography{references, zotero_references}

\end{document}


\preprint{APS/123-QED}

\title{Supplemental material for "Electronic and magnetic properties of light rare-earth cubic Laves compounds derived from XMCD experiments"}

\author{Vilde G. S. Lunde}\thanks{Contact author: vilde.lunde@ife.no}
\affiliation{Department for Hydrogen Technology, Institute for Energy Technology, PO Box 40, NO-2027, Kjeller, Norway}
\author{Benedicte S. Ofstad}
\affiliation{Department for Hydrogen Technology, Institute for Energy Technology, PO Box 40, NO-2027, Kjeller, Norway}
\author{{\O}ystein S. Fjellv{\aa}g}
\affiliation{Department for Hydrogen Technology, Institute for Energy Technology, PO Box 40, NO-2027, Kjeller, Norway}
\author{Philippe Ohresser}
\affiliation{Synchrotron SOLEIL, L'Orme des Merisiers, 91190, Saint-Aubin, France}
\author{Anja O. Sj{\aa}stad}
\affiliation{Chemistry Department and Center for Material Science and Nanotechnology, University of Oslo, NO-0315, Norway}
\author{Bj{\o}rn C. Hauback}
\affiliation{Department for Hydrogen Technology, Institute for Energy Technology, PO Box 40, NO-2027, Kjeller, Norway}
\author{Christoph Frommen}\thanks{Contact author: christoph.frommen@ife.no}
\affiliation{Department for Hydrogen Technology, Institute for Energy Technology, PO Box 40, NO-2027, Kjeller, Norway}

\renewcommand{\thefigure}{A\arabic{figure}}
\setcounter{figure}{0}
\renewcommand{\thetable}{A\arabic{table}}
\setcounter{table}{0}
\renewcommand{\theequation}{A\arabic{equation}}
\setcounter{equation}{0}

\date{\today}

\begin{abstract}
This is the supplemental material. Section~\ref{sec:multiplet} provides a detailed description of the multiplet theory calculations performed, Section~\ref{sec:XRD} contains refined XRD patterns with lattice constants in Section~\ref{sec:XRD}, Section~\ref{sec:XAS_XMCD} presents XAS and XMCD experimental and simulated spectra. Results from the sum rule calculations of XMCD data are in Section~\ref{sec:sum_rules}.

\end{abstract}

\maketitle

\section{Multiplet Theory Methodology}
\label{sec:multiplet}

\textsc{quanty} was used for full-multiplet calculations of the crystal electric field. Single-ion Hamiltonian, on-site Coulomb repulsion $\mathcal{H}_\text{U}$, in addition to the usual spin-orbit coupling $\mathcal{H}_\text{SOC}$, and a crystal electric field potential $\mathcal{H}_\text{CEF}$ Hamiltonian:

\begin{align}
\mathcal{H} &= \mathcal{H}_\text{U} + \mathcal{H}_\text{SOC} + \mathcal{H}_\text{CEF},   \\ 
&=  \mathcal{H}_\text{U} + \zeta \mathbf{L} \cdot \mathbf{S} + \sum_{l\in\{2,4,6\}}  \sum_{m=-l}^l A_l^m C_l^m(\theta,\phi)  ,
\label{ham}
\end{align}

\noindent with the coupling constant $\zeta$ of the orbital angular momentum $\mathbf{L}$, and spin angular momentum $\mathbf{S}$ determined experimentally. $\mathcal{H}_\text{CEF}$ is expanded over spherical harmonics $Y_l^m(\theta)$ with expansion coefficients $A_l^m$. We sum over the even orbital angular momentum quantum number $l$.

The crystal electric field was modeled assuming cubic point group symmetry, with the relevant expansion coefficients $A^0_4$ and $A^0_6$ treated as adjustable parameters for Nd and Pr as a cubic crystal field requires $A_4^{\pm4} = \sqrt{\frac{5}{14}}A_4^0$ and $A_6^{\pm 4} = -\sqrt{\frac{7}{2}}  A_6^0$. 
On the other hand, Ce has $J=\sfrac{5}{2}$ and the expansion coefficients $A_m^l$ can only act on a multiplet with $m$ larger than $2J$. 
Thus, is $A^0_4$ the only free parameter for Ce.  
The correct crystal field parameters were identified by restricting the model to the XMCD spectra and the orbital moment, as determined by the orbital sum rules. 
As one or two parameters describe the crystal field, it is possible to find the global minimum.
XAS and XMCD spectra were generated with a field of 5~T and a temperature of 4.2~K, as was done in the experiments.

\newpage 
\section{X-Ray Diffraction}
\label{sec:XRD}

\begin{figure*}[h!]
    \centering
    \includegraphics[width = 0.9\textwidth]{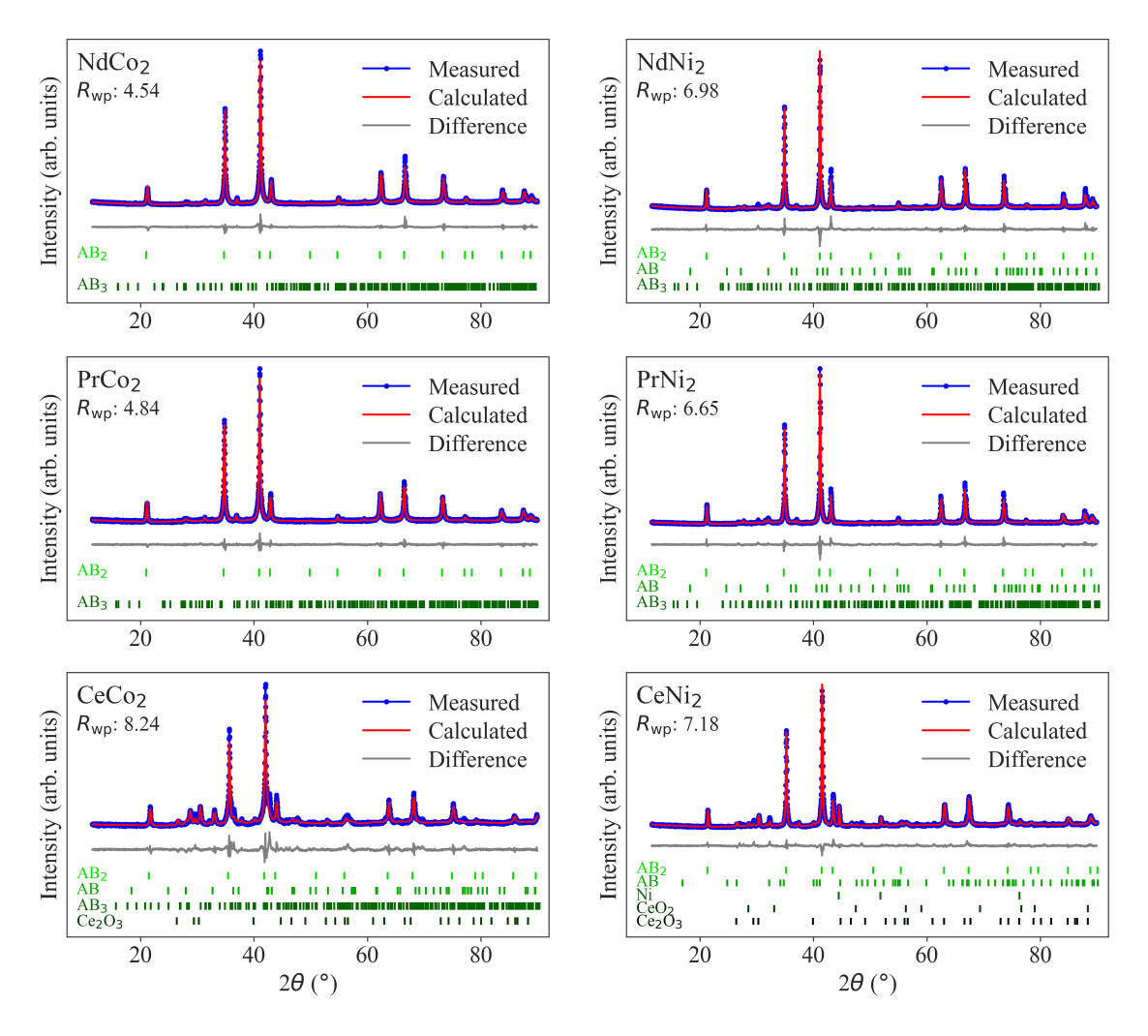}
    \caption{Rietveld refinement of the XRD patterns of the reference samples, measured at room temperature, with a wavelength of 1.54060~{\AA}. K$\beta$ was added due to imperfect K$\beta$ elimination by the detector.}
    \label{fig:Refinement}
\end{figure*}

\begin{figure*}[h!]
    \centering
    \includegraphics[width = 0.55\textwidth]{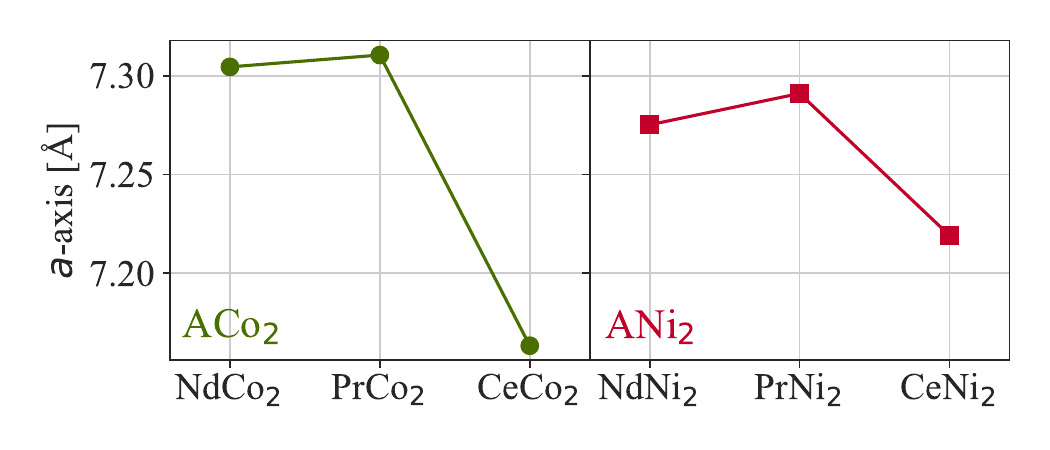}
    \caption{Lattice constants for the \ce{AB2} primary phase of the reference samples, based on Rietveld refinement of XRD-data, shown in Figure~\ref{fig:Refinement}. Error bars are within the size of the markers.}
    \label{fig:Lattice_Constants}
\end{figure*}

\newpage
\section{Density of States}
\label{sec:DOS}

\begin{figure*}[h!]
    \centering
    \includegraphics[width = 0.65\textwidth]{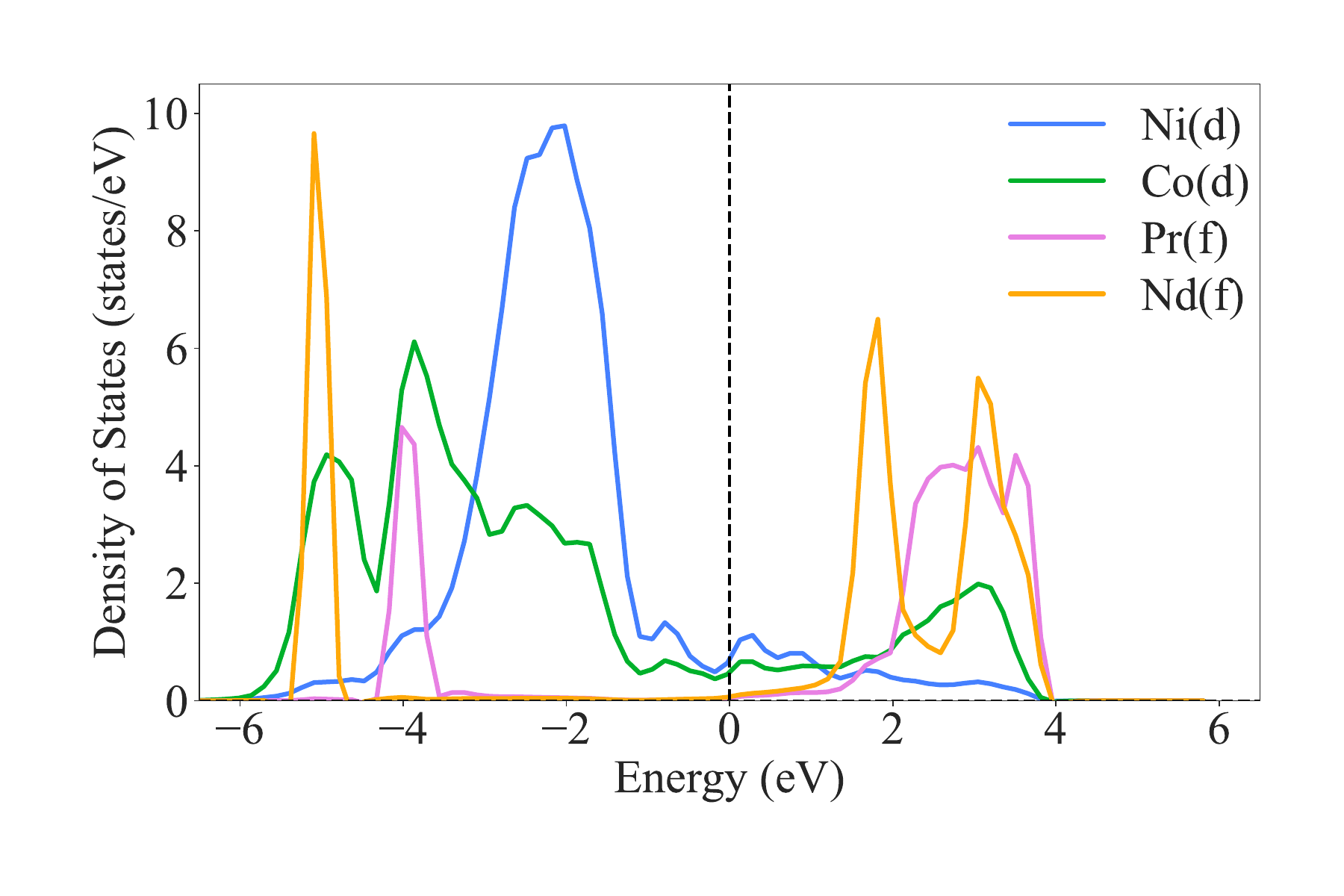}
    \caption{Density of states for \ce{Nd_{0.50}Pr_{0.50}CoNi}, generated using DFT. The Fermi energy is marked with a vertical dashed line.}
    \label{fig:NdPrCoNi_dos}
\end{figure*}

\newpage
\section{XAS and XMCD Spectra}
\label{sec:XAS_XMCD}

\begin{figure*}[h!]
  \centering
  \subfloat[][]{\includegraphics[width = 0.5\textwidth]{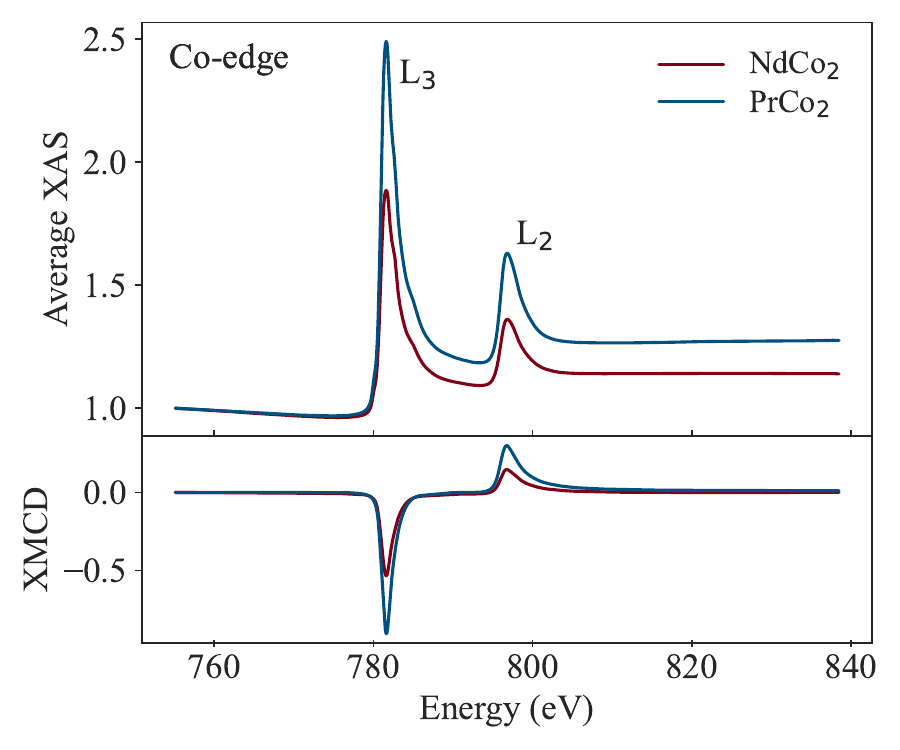} \label{fig:XMCD_Binaries_Co}} 
  \subfloat[][]{\includegraphics[width = 0.5\textwidth]{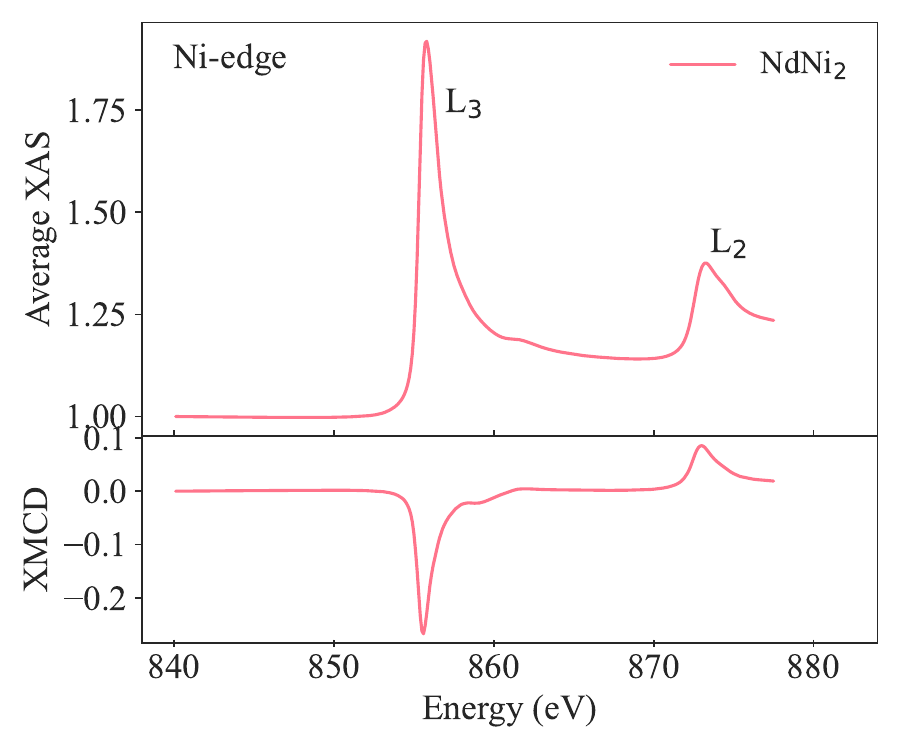} \label{fig:XMCD_Binaries_Ni}} \\
  \subfloat[][]{\includegraphics[width = 0.5\textwidth]{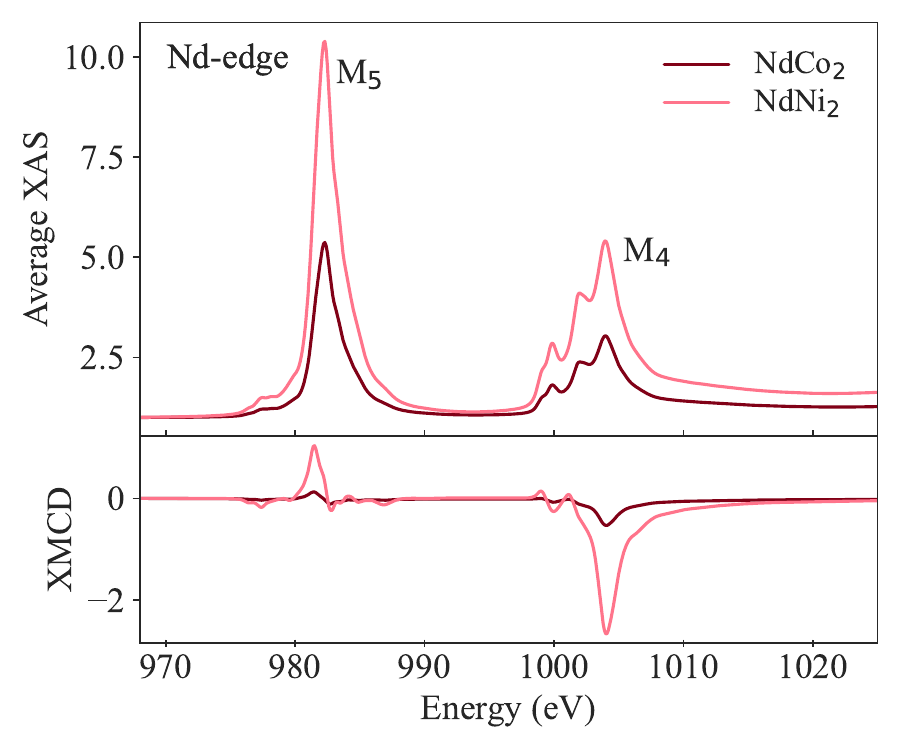} \label{fig:XMCD_Binaries_Nd}} 
  \subfloat[][]{\includegraphics[width = 0.5\textwidth]{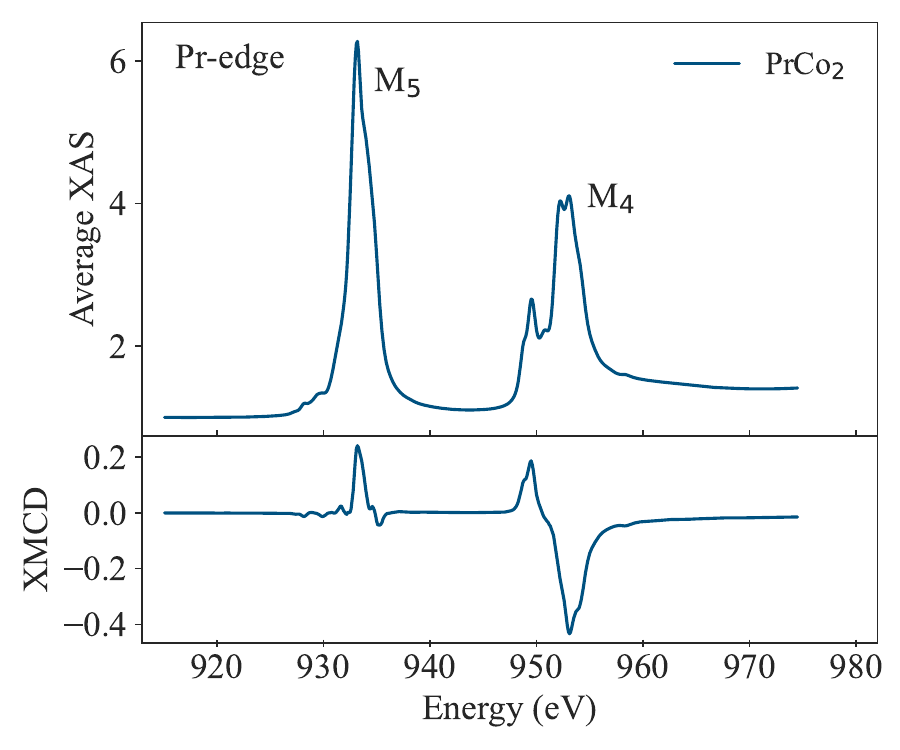} \label{fig:XMCD_Binaries_Pr}} \\
  \caption{XAS and XMCD for the binaries in a ferromagnetic state. Average XAS shows the average of the right and left polarized x-ray absorption, while XMCD shows the difference.} \label{fig:XMCD_Binaries}
\end{figure*}

\newpage
\section{Sum Rules}
\label{sec:sum_rules}

\begin{table*}[h!]
\centering 
\caption{\label{tab:XMCD} Orbital and spin magnetic moments at \SI{4.2}{K} and \SI{5}{T} calculated using the magneto-optical sum rules. $n_h$ was calculated using DFT. Values for binaries in a ferromagnetic state are included. Values calculated using the spin sum rule for the rare earth elements are written in cursive to underline their invalidity.}
\begin{ruledtabular}
\begin{tabular}{l l c c c c c}
\\[-3mm]
Composition & Atom & $n_h$ & $\mu_\text{L}$ & $\mu_\text{S}$ & $\mu_\text{tot}$ & $\mu_\text{L}$/$\mu_\text{S}$ \\
  &  &    & ($\mu_\text{B}$/atom) & ($\mu_\text{B}$/atom) & ($\mu_\text{B}$/atom) &   \\[1mm] 
\hline \\[-2mm]
\ce{NdCo_2}                 & Co & 1.56 & 0.23 & 1.05 & 1.28 & 0.22 \\
\ce{PrCo_2}                 & Co & 1.56 & 0.10 & 1.22 & 1.32 & 0.08 \\
\ce{NdCoNi}                 & Co & 1.56 & 0.15 & 1.18 & 1.32 & 0.12 \\
\ce{Nd_{0.75}Pr_{0.25}CoNi} & Co & 1.56 & 0.07 & 1.11 & 1.18 & 0.06 \\
\ce{Nd_{0.50}Pr_{0.50}CoNi} & Co & 1.56 & 0.09 & 1.32 & 1.41 & 0.07 \\
\ce{Nd_{0.25}Pr_{0.75}CoNi} & Co & 1.56 & 0.12 & 1.17 & 1.29 & 0.10 \\
\ce{PrCoNi}                 & Co & 1.56 & 0.14 & 1.28 & 1.42 & 0.11 \\
\ce{Ce_{0.25}Pr_{0.75}CoNi} & Co & 1.56 & 0.16 & 1.13 & 1.29 & 0.14 \\[1mm] 

\hline \\[-3mm]

\ce{NdNi_2}                 & Ni & 1.07 & 0.07 & 0.34 & 0.41 & 0.20 \\
\ce{NdCoNi}                 & Ni & 1.07 & 0.06 & 0.54 & 0.60 & 0.11 \\
\ce{PrCoNi}                 & Ni & 1.07 & 0.05 & 0.56 & 0.61 & 0.09 \\[1mm]

\hline \\[-3mm]

\ce{NdCo_2}                 & Nd & 11.00 & 1.28 & \textit{1.58*} & \textit{-0.30*} & \textit{0.81*} \\
\ce{NdNi_2}                 & Nd & 11.00 & 1.94 & \textit{3.24*} & \textit{-1.31*} & \textit{0.60*} \\
\ce{NdCoNi}                 & Nd & 11.00 & 1.69 & \textit{2.80*} & \textit{-1.11*} & \textit{0.60*} \\
\ce{Nd_{0.75}Pr_{0.25}CoNi} & Nd & 11.00 & 1.46 & \textit{3.01*} & \textit{-1.55*} & \textit{0.49*} \\
\ce{Nd_{0.50}Pr_{0.50}CoNi} & Nd & 11.00 & 1.68 & \textit{3.27*} & \textit{-1.59*} & \textit{0.51*} \\
\ce{Nd_{0.25}Pr_{0.75}CoNi} & Nd & 11.00 & 1.19 & \textit{2.95*} & \textit{-1.75*} & \textit{0.41*} \\[1mm] 

\hline \\[-3mm]

\ce{PrCo_2}                 & Pr & 12.00 & 0.25 & \textit{0.72*} & \textit{-0.47*} & \textit{0.35*} \\
\ce{Nd_{0.75}Pr_{0.25}CoNi} & Pr & 12.00 & 0.31 & \textit{1.32*} & \textit{-1.00*} & \textit{0.24*} \\
\ce{Nd_{0.50}Pr_{0.50}CoNi} & Pr & 12.00 & 0.54 & \textit{1.35*} & \textit{-0.81*} & \textit{0.40*} \\
\ce{Nd_{0.25}Pr_{0.75}CoNi} & Pr & 12.00 & 0.55 & \textit{1.25*} & \textit{-0.70*} & \textit{0.44*} \\
\ce{PrCoNi}                 & Pr & 12.00 & 0.49 & \textit{1.32*} & \textit{-0.83*} & \textit{0.37*} \\
\ce{Ce_{0.25}Pr_{0.75}CoNi} & Pr & 12.00 & 0.55 & \textit{1.25*} & \textit{-0.70*} & \textit{0.44*} \\[1mm] 

\hline \\[-3mm]

\ce{Ce_{0.25}Pr_{0.75}CoNi} & Ce & 13.00 & 0.53 & \textit{0.68*} & \textit{-0.15*} & \textit{0.77*} \\[1mm] 

\end{tabular}
*Calculated using the invalid spin sum rule for 4$f$ elements.
\end{ruledtabular}
\end{table*}

\begin{table*}[h!]
\centering 
\caption{\label{tab:XMCD2} Orbital and spin magnetic moments at \SI{4.2}{K} and \SI{5}{T} calculated using multiplet theory. Parameters were chosen so that $\mu_\text{L}$ was the same as the experimental, and these parameters were used to calculate $\mu_\text{S}$.}
\begin{ruledtabular}
\begin{tabular}{l l c c c c c}
\\[-3mm]
Composition & Atom & $n_h$ & $\mu_\text{L}$ & $\mu_\text{S}$ & $\mu_\text{tot}$ & $\mu_\text{L}$/$\mu_\text{S}$ \\
  &  &    & ($\mu_\text{B}$/atom) & ($\mu_\text{B}$/atom) & ($\mu_\text{B}$/atom) &   \\[1mm] 
\hline \\[-3mm]
\ce{NdCo_2}                 & Nd & 11.00 & 1.28 & 0.32 & 0.96 & 4 \\
\ce{NdNi_2}                 & Nd & 11.00 & 1.94 & 0.48 & 1.45 & 4 \\
\ce{NdCoNi}                 & Nd & 11.00 & 1.69 & 0.42 & 1.26 & 4 \\
\ce{Nd_{0.75}Pr_{0.25}CoNi} & Nd & 11.00 & 1.46 & 0.36 & 1.09 & 4 \\
\ce{Nd_{0.50}Pr_{0.50}CoNi} & Nd & 11.00 & 1.68 & 0.42 & 1.26 & 4 \\
\ce{Nd_{0.25}Pr_{0.75}CoNi} & Nd & 11.00 & 1.19 & 0.30 & 0.89 & 4 \\[1mm] 
\hline \\[-3mm]
\ce{PrCo_2}                 & Pr & 12.00 & 0.25 & 0.06 & 0.19 & 4 \\
\ce{Nd_{0.75}Pr_{0.25}CoNi} & Pr & 12.00 & 0.31 & 0.08 & 0.24 & 4 \\
\ce{Nd_{0.50}Pr_{0.50}CoNi} & Pr & 12.00 & 0.54 & 0.13 & 0.40 & 4 \\
\ce{Nd_{0.25}Pr_{0.75}CoNi} & Pr & 12.00 & 0.55 & 0.14 & 0.41 & 4 \\
\ce{PrCoNi}                 & Pr & 12.00 & 0.49 & 0.12 & 0.37 & 4 \\
\ce{Ce_{0.25}Pr_{0.75}CoNi} & Pr & 12.00 & 0.55 & 0.14 & 0.41 & 4 \\[1mm] 
\hline \\[-3mm]
\ce{Ce_{0.25}Pr_{0.75}CoNi} & Ce & 13.00 & 0.53 & 0.13 & 0.40 & 4 \\[1mm]
\end{tabular}
\end{ruledtabular}
\end{table*}
